\documentclass[11pt,preprint]{aastex}
\usepackage{epsfig,lscape}
\slugcomment{submitted to ApJ}

\def\la{\mathrel{\mathpalette\fun <}}

\def\fun#1#2{\lower3.6pt\vbox{\baselineskip0pt\lineskip.9pt
        \ialign{$\mathsurround=0pt#1\hfill##\hfil$\crcr#2\crcr\sim\crcr}}}
        
\tighten

\begin{document}

\title{Wavelet Band Powers of the Primordial Power Spectrum from CMB Data}
\author{Pia~Mukherjee, Yun~Wang}
\affil{Department of Physics \& Astronomy, Univ. of Oklahoma,
                 440 W Brooks St., Norman, OK 73019\\
                 email: pia@nhn.ou.edu, wang@nhn.ou.edu\\
                 (\today)}

\begin{abstract}

Measuring the primordial matter power spectrum is our 
primary means of probing unknown physics in the very early universe.
We allow the primordial power spectrum to be an arbitrary function, 
and parametrize it in terms of wavelet band powers. 
Current cosmological data correspond to 11 such wavelet bands. We 
derive constraints on these band powers as well as $H_0$, 
$\Omega_b h^2$ and $\Omega_m h^2$ from 
current Cosmic Microwave Background Anisotropy (CMB) data 
using the Markov Chain Monte Carlo (MCMC) technique.  Our results 
indicate a feature in the primordial power spectrum at 
$ 0.008 \la k/(h\,\mbox{Mpc}^{-1} ) \la 0.1$. MAP and Planck data should 
allow us to put tighter constraints on the primordial power spectrum.

\end{abstract}

\section{Introduction}

As a result of recent cosmological data,
inflation \citep{Guth81,Albrecht82,Gott82,Linde83}
has become increasingly well established 
as the plausible solution to the problems of standard cosmology
(\cite{Kolb&Turner}; see \cite{peeblesratra} for a recent review).
The primordial power spectrum is our primary window into
unknown physics during inflation \citep{Wang99,Chung00,Enqvist00,Lyth02}.
It is of critical importance that we try to extract the primordial power
spectrum, $P_{in}(k)$, from observational data without assuming specific forms
for it.

Cosmological parameters are being measured to impressive precision 
with the help of recent CMB and large scale structure (LSS) data. 
The parameter constraints thus deduced 
are however sensitive to assumptions regarding the 
power spectrum of primordial density perturbations (\cite{Kinney01, Souradeep98}). 
The primordial power 
spectrum is often assumed to be a power law, which represents
many inflationary models (for example, see
\cite{Linde83,naturalinf,extendedinf}).
With such a parametrization the 
primordial power spectrum has been found to be scale invariant to a very 
good approximation, and its amplitude constrained 
(see for example \cite{gorski94,LB02}).
However, there are many inflation models that predict primordial power 
spectra which cannot be parametrized by a simple power law 
(for example, 
\cite{Holman91ab,Wang94,Randall96,Adams97,Les97,Les99}).
These can represent unusual physics in the very early universe.
For example, inflation might occur in multiple stages
in effective theories with two scalar fields \citep{Holman91ab}, 
or in a succession of short bursts due to
symmetry breaking during an era of inflation in 
supergravity models \citep{Adams97}.

With the quality of data 
improving, more attention is being paid to the nature of the primordial 
perturbations (for example, see \cite{Covi02,LeachLiddle02}).
As more observational data become available,
they increase our ability to probe the 
primordial power spectrum, $P_{in}(k)$, as an arbitrary function of scale. 
A model independent 
determination of the $P_{in}(k)$ could uniquely 
probe physics of the very early universe, test what we have  
assumed about early universe physics, and provide powerful 
constraints on inflationary models.  
\cite{Wang99} explored how this can be done 
with the CMB data from MAP\footnote{http://map.gsfc.nasa.gov/}
 and LSS data from SDSS, using a 
piecewise constant function for $P_{in}(k)$. 
\cite{WangMathews02} used the CMB data from Boomerang, Maxima, and
DASI to place constraints on $P_{in}(k)$ using 
linear interpolation to approximate the function between several 
$k$ values equally spaced in log$k$. 

Here we employ wavelets in a model independent parametrization 
of the primordial power spectrum and obtain constraints from 
current CMB data. We use only CMB data here, but LSS
(for example, see \cite{Hamilton02,Percival02,Bahcall02}),
and CMB polarization (for example, see \cite{Kovac02}) data 
can all be added to help break or reduce 
degeneracies between different cosmological parameters and 
help better constrain $P_{in}(k)$.
 
We describe the wavelet parametrization of the primordial power
spectrum in Sec.2 (see Appendix A for further discussion). 
In Sec.3, we discuss the techniques that we
have used to optimize our method. Sec.4 contains our results.
Sec.5 contains a summary and discussions.

\section{Wavelet Band Powers of the Primordial Power Spectrum}

Assuming that the primordial matter density fluctuations form a
 homogeneous 
Gaussian random field, the wavelet band powers of
the primordial power spectrum $P_{in}(k)$, can be written as 
(see Eq.(\ref{impeqn1}))
\begin{equation}
P_j = \frac{1}{2^j} \int_{-\infty}^{\infty}dk\, \left| \hat{\psi}
\left(\frac{k}{2^j}\right) \right|^2 P_{in}(k),
\label{eq:Pj}
\end{equation} 
where $\hat{\psi}$ is the Fourier transform of
the wavelet ${\psi}$ (see Appendix A).
$P_j$'s represent a scale-by-scale band averaged Fourier power spectrum,
where the banding is not arbitrary but well defined, and naturally adaptive 
(as wavelets by construction afford better $k$ resolution at smaller $k$). 
The $P_j$'s are mutually uncorrelated by construction 
(see Appendix A and \cite{fangfeng} for further discussion of these issues).

To illustrate, Figure 1 shows a primordial power spectrum with features (solid line),
similar to that discussed by \cite{Elgaroy02}.
The points are the wavelet band powers $P_j$, 
computed using Eq.(\ref{eq:Pj}).
These $P_j$'s can be thought of as being measured exactly from ideal data.
Clearly, the wavelet band powers of the primordial power spectrum 
are excellent approximations
of the primordial power spectrum at the central $k$ values of the
wavelet window functions, $\left| \hat{\psi}\left(\frac{k}{2^j}
\right)\right|^2$ (dotted curves in Fig.1). Since the window
 functions are the modulus
 squared of the Fourier transform of dilations, by factors of 2, of the 
basic wavelet (see Appendix A), the peaks of these window functions are separated by
 factors of 2 in $k$. While we can add more bands to lower $k$,
 and 3 more bands at higher $k$ that go up to $2\pi$ in $k$, and 
reconstruct the amplitude of the power in these additional bands, 
 the locations of the bands or their widths cannot be changed.
 This is why in this scheme 
the banding is not arbitrary.

\begin{figure}[p]
\psfig{file=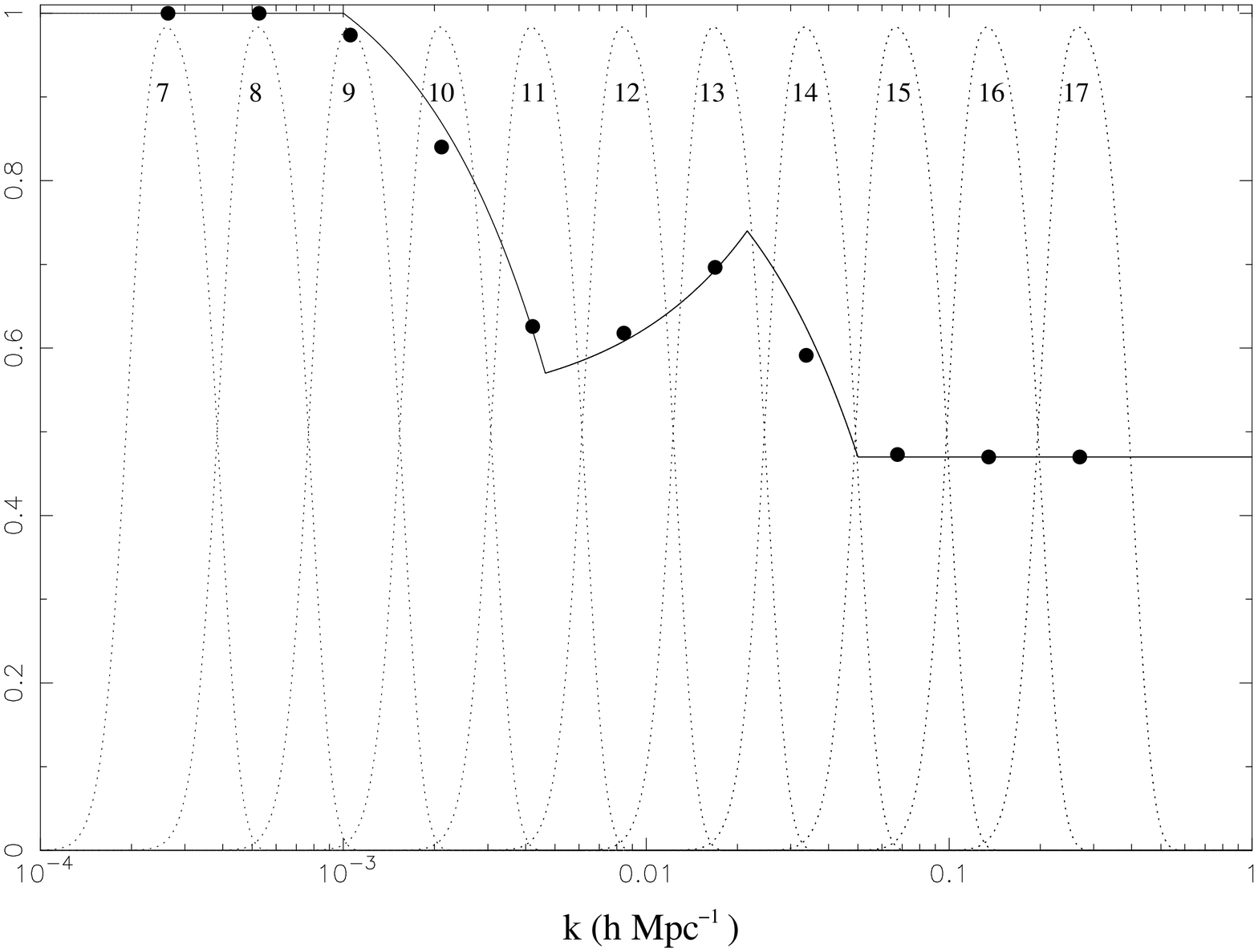,height=2.5in,angle=0}
\caption{An example of a primordial power spectrum (solid line), 
and its wavelet band powers $P_j$'s ($j$=7,17; points), together with the
 corresponding wavelet window functions (dotted lines).
The y axis is in arbitrary units.}
\end{figure}

Hence we parametrize the primordial power spectrum
$P_{in}(k)$ as follows [see Eq.(\ref{impeqn2})]
\begin{equation}
P_{in}(k) = \sum_{j} P_j \left| \hat{\psi}\left(\frac{k}{2^j}
\right)\right|^2.
\label{eq:Pin(k)}
\end{equation}
The wavelet band powers $P_j$ can be estimated from data that depend
on the primordial power spectrum $P_{in}(k)$, for example, CMB
and LSS data. Here we use only CMB data and estimate cosmological parameters 
together with the wavelet band powers $P_j$, which represent a measurement of 
the primordial
power spectrum at the central $k$ values of the
wavelet window functions. 
$P_j$'s of unity corresponds to the default power of $A_s^2=2 \times 10^{-9}$,
 and if all $P_j$'s are set to unity we recover 
exactly the $C_l$ spectrum that results from a scale invariant 
primordial power spectrum 
$P_s(k)=A_s^2 \left( \frac{k}{k_0} \right)^{n_s-1}$. 

Figure 2 shows how each wavelet band power window function,
$\left| \hat{\psi}\left(\frac{k}{2^j}\right)\right|^2$ 
[shown in Figure 1], maps on to a window function in the CMB angular power
spectrum multipole number $l$ space, for one set of cosmological parameters. 
These 
window functions have been numbered 7 to 17, which correspond to a range of 
$0.0001  \la k /(h\, \mbox{Mpc}^{-1}) \la 0.5$. 
The 11 band powers, $P_{7}$ through $P_{17}$ (from small to large $k$), are 
sufficient to constrain 
the primordial power spectrum from current and near future cosmological data. 
CMB data, up to an $l_{max}$ of 1500, are insensitive to band powers $P_1$ to $P_6$, 
which correspond to smaller 
$k$ and have been set equal to $P_7$, and to band powers higher than $P_{17}$, 
which correspond to larger $k$ and have been set equal to $P_{17}$. 
The solid curve is the CMB angular power spectrum $C_l$ that 
includes contributions from all the $P_j$'s [set to unity in this Figure]. 
Given that the CMB data are in the form of band 
powers in $C_l$s, and each multipole $l$ maps on to a range in $k$ 
(for example, see Figure 4 of \cite{Tegmark02}) with dependence 
on cosmological parameters,
we see that in constraining $P_j$ using 
cosmological data some correlations between the $P_j$'s will be inevitably 
introduced. 
The $P_j$'s are also correlated somewhat with the cosmological parameters 
in ways that can be understood from Figure 2.

For a comparison of different banding methods, instead of wavelet band powers 
we can parametrize the
 primordial power spectrum as a continuous and arbitrary function determined
 by its amplitude at several wavenumbers, equally spaced in log$(k)$, using
linear interpolation to approximate the function at intermediate wavenumbers.
 This is the linear interpolation binning method of \cite{WangMathews02}.  
 In this work, we determine amplitudes
 at 11 wavenumbers corresponding to the central $k$ values of the wavelet 
band powers.  Figure 3 shows how each of these ``binned'' amplitudes 
map onto the CMB angular power spectrum multipole number $l$.
The solid curve is the CMB angular power spectrum $C_l$
with contributions from all the bins [all bin amplitudes set to 1]. 
The power in each bin, denoted here by $a_j$, is correlated with the power 
in neighboring bins from the start, and ultimately the resulting 
correlations will be different from the wavelet banding method. 
The error bars on the 
estimates will also be correlated. See \cite{WangMathews02} for further 
details about this method.

\begin{figure}[p]
\psfig{file=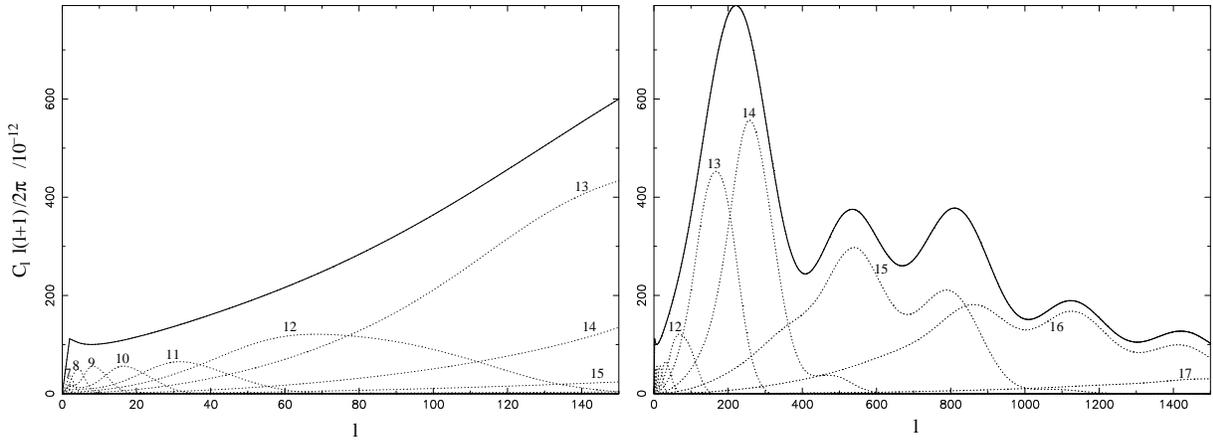,height=2.3in,angle=0}
\caption{Mapping of the wavelet window functions of Figure 1 into
window functions in the CMB multipole $l$ space (dotted curves). 
The solid curve is the $C_l$ spectrum that is the sum of contributions from
all the wavelet bands (all the band powers, $P_j$, are set to unity here 
for illustration).}
\end{figure}

\begin{figure}[p]
\psfig{file=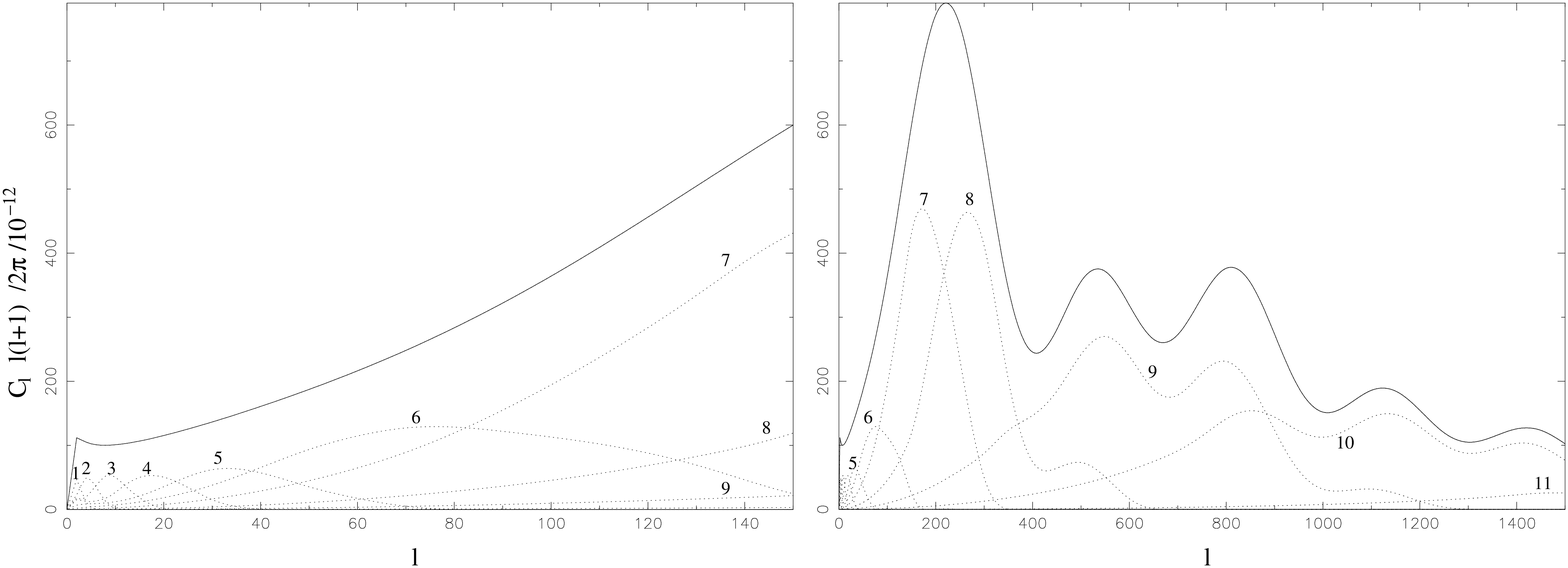,height=2.3in,angle=0}
\caption{Mapping of an arbitrary number of bins, here chosen to be 11, 
to correspond exactly to the central k values of the 11 wavelet bands 
discussed above, into window functions in the CMB multipole $l$ space 
(dotted curves).
 The solid line is the $C_l$ spectrum that includes contributions from all the
bins (all the bin amplitudes are set to unity here for illustration).}
\end{figure}

\section{Method}

\subsection{Optimization Techniques: Wavelet Projections and Markov Chain Monte Carlo}

The usual approach to deriving cosmological constraints from
CMB data is to grid over all the parameters being estimated
and compute the likelihood at each point. 
Here, besides the 11 wavelet band powers ($P_j$'s), we constrain 
$H_0$, $\Omega_b h^2$ and $\Omega_c h^2$, 
assuming a flat universe with a cosmological constant $\Lambda$, and 
ignoring reionization and tensor modes. 
This results in 14 parameters.
In order to place the most accurate and reliable constraint
on the primordial power spectrum, we have chosen to compute
the theoretical CMB angular power spectra at the accuracy 
of CMBFAST \citep{SeljakZ96} and CAMB \citep{Lewis00}.
Our analysis would have been extremely
time-consuming even on supercomputers.
Fortunately, we are able to use two techniques which have made it possible 
for the analysis to be completed in a timely fashion.

First is the wavelet projections of the CMB angular power spectrum, $C_l$'s.
Using Eq. (2), we expand the CMB angular 
power spectrum as follows:
\begin{eqnarray}
C_l(P_j, \mbox{\bf s}) &=&(4\pi)^2 \int \frac{dk}{k} P_{in}(k) 
\left|\Delta_{Tl}(k, \tau=\tau_0)\right|^2 \nonumber \\
&=& \sum_{j} P_j \int \frac{dk}{k} \left|\hat{\psi}
\left(\frac{k}{2^j}\right) \right|^2 \left|\Delta_{Tl}(k, \tau=\tau_0)\right|^2\equiv 
\sum_j P_j C_l^j (\mbox{\bf s}),
\label{Clwaveletproj}
\end{eqnarray}
 where the cosmological model dependent transfer function 
$\Delta_{Tl}(k,\tau=\tau_0)$ is an integral over
 conformal time $\tau$ of the sources which generate CMB fluctuations,
 $\tau_0$ being the conformal time today,
 and $\mbox{\bf s}$ represents the cosmological parameters other than
the $P_j$'s. We use CAMB\footnote{http://camb.info/} to compute the 
CMB angular power spectrum,
 in a form such that for given cosmological 
parameters other than the $P_j$'s, the $C_l^j(\mbox{\bf s})$ are computed,
so that there is no need to call CAMB when we vary the $P_j$'s.
This results in significant computational speed up.  
We expect this technique to be essential in constraining the 
primordial power spectrum with sufficient detail and speed. 

The second technique is the Markov Chain Monte Carlo (MCMC) technique.
The large number of parameters being varied here 
necessitates the use of this technique in the likelihood analysis.  
At its best, the MCMC method scales approximately linearly with the number of 
parameters. The method samples from the full posterior distribution of the
 parameters, and from these samples the marginalized posterior distributions
 of the parameters can be estimated.  See \cite{neil} for a review, and
 \cite{hannestad, Knox01, LB02, Rubino-Martin02} for applications of
 this method to CMB analysis. 

\subsection{Recipe for the Analysis}

The steps followed in the analysis are as follows:

\noindent
(i) Establish the multi-dimensional parameter space that span all the parameters
to be estimated from data. In our case, the total number of parameters is 14,
including the wavelet band powers $P_j$'s.\\
\noindent
(ii) Start at some point in our 14d parameter space, which corresponds to
an initial set of parameters. 
Note that the MCMC chain soon loses memory of this starting point and it can be 
verified that the result is independent of it. \\
\noindent
(iii) Find the likelihood of this set of 
parameter values given the data: 
\begin{equation}
-2ln(L) = \chi^2 \equiv \sum_{b,b'} 
\left[ln(C_b + x_b)-ln(C_b^{th}+ x_b)\right] G_{b,b'} \left[ln(C_{b'} + x_{b'})-
ln(C_{b'}^{th}+ x_{b'})\right],
\end{equation}
 where $C_b$ and $C_b^{th}$ are the experimental
 and theoretically estimated band powers respectively, $G_{b,b'}$ is
the corresponding
 band-band correlation matrix and $x_b$ are the offset parameters
 in the offset-lognormal ansatz of Bond et al. (1998), or 0 for experimental
 results computed without this ansatz.  For this purpose, we calculate 
the $C_l$ spectrum
for the given set of parameters, estimate the expected experimental
 band powers 
(for the given set of parameters) using the band power window functions 
available for the different CMB experiments, and compare these estimates 
to the 
measured band powers, taking into account the instrumental noise. We use 
offset-lognormal band-powers whenever available, and 
analytically marginalize over known beam and calibration uncertainties 
for each experiment, as described in Bridle et al. (2002). \\
\noindent
(iv) Take a random step in parameter space. The probability distribution of the step is taken to be Gaussian in each direction with an $rms$ that is neither too small (or the chain will exhibit poor mixing) nor too large (or the chain efficiency will be poor). Ensure that the new point lies within the wide priors that define the likelihood region that one chooses to explore (our priors are defined in the next section). Calculate the likelihood at 
this new point (which corresponds to a new set of parameters). \\
\noindent
(v) Use the Metropolis-Hastings algorithm (an MCMC sampler 
based on the Metropolis-Hastings
algorithm has been made available in the software package CosmoMC 
\citep{LB02}) to determine whether to accept this point or to discard it 
and propose another point in the parameter space.\\
\noindent
(vi) Repeat steps (iii) to (v) until a large number of samples from the posterior 
distribution of these parameters have been chosen. \\
\noindent
(vii) Find the 1d marginalized parameter distributions 
and confidence limits for each parameter. \\
(viii) Check for convergence by running several chains 
starting at different initial values giving the same final distributions, 
and a couple of longer chains to ensure good sampling.

It would be interesting to include reionization and curvature in the analysis, 
and obtain additional constraints from different kinds of LSS data and ultimately 
also CMB polarization. We defer such extensions to the near future.

\section{Results and Discussion}

We use most of the recent high precision
data for which experimental details are publicly available.
These are the latest Boomerang \citep{boom}, Maxima \citep{maxima}, 
DASI \citep{dasi}, VSA \citep{vsa}, CBI \citep{cbi}, ACBAR \citep{acbar} and Archeops
 \citep{archeops}, together with COBE
 \citep{Smoot92}. 
We estimate the wavelet band powers of the primordial power spectrum,
as well as $H_0$, $\Omega_b h^2$ and $\Omega_c h^2$.

All results shown in this paper are for the wavelet Daubachies 20
[see Appendix A].

Allowing all 14 parameters to vary, the results obtained 
are shown in Figure 4. 
We have used a weak prior of the age of the universe $t_0 > 10$ Gyrs. All the 
$P_j$'s are consistent with unity at $\sim 1.5\sigma$. However,
there is some indication of a feature in $P_{in}(k)$, though
at low significance.
Figure 4 shows that current CMB data seem to favor a dip at a $k$ of
 $\sim 0.01\,h\,$Mpc$^{-1}$ 
(from $P_{12}$) and excess power $k$ of $\sim 0.02$ to 0.03$\,h\,$Mpc$^{-1}$
 (from $P_{13}$ 
and $P_{14}$), followed by another small dip from $P_{16}$ 
at $k$ of $\sim 0.1\,h\,$Mpc$^{-1}$. 
The cosmological parameters are constrained to   
$h=0.56\pm0.09$, 
$\Omega_b h^2=0.020\pm 0.005$ and $\Omega_c h^2 = 0.161\pm 0.028$.\footnote{In
 this paper we quote the mean of the derived 1d distributions, instead of the
 maximum, for the constrained parameters, following \cite{LB02}. As discussed
 in \cite{LB02}, the MCMC samples from the posterior do not provide accurate
 estimates of parameter best-fit values, because in higher dimensions the 
best-fit region typically has a much higher likelihood than the mean but
 occupies a minuscule fraction of parameter space.  Our main results are 
not affected by this.},\footnote{This corresponds to 
$\Omega_b=0.062 [0.048, 0.079]$ and $\Omega_c=0.51 [0.42, 0.60]$.  Note the 
degeneracies between the cosmological parameters considered here.  Due to 
the geometrical degeneracy, the location of the first acoustic peak nails 
the curvature of the universe (here taken to be zero) but a 
degeneracy between $\Omega_m$ and $\Omega_{\Lambda}$, or equivalently 
$\Omega_m$ and $h$ remains.  The height of the first peak is degenerate for
 $\Omega_b h^2$ and $\Omega_c h^2$. Precise determination of the second 
 and third peak heights can help determine $\Omega_b h^2$ and thus ease 
degeneracies
(for example, see \cite{Efs99,Hu01,White02}).} 
Note that the derived value of the Hubble constant $h$ is lower than 
the value of $h$
derived assuming a scale-invariant primordial power spectrum (i.e., setting
all the $P_j$'s to the same constant value). 
Assuming scale-invariance gives $h=0.71\pm 0.06$, 
$\Omega_b h^2=0.023\pm 0.001$, $\Omega_c h^2=0.122\pm 0.016$ and all 
$P_j$'s$=0.90\pm 0.06$.
A chi square analysis indicates that a model with the 
$P_j$'s estimated from data (i.e., allowing $P_{in}(k)$ to be
a free function) and the best fit scale 
invariant model (setting all $P_j$'s equal to a constant)
fare about the same. 
For a low $h$, the derived $\Omega_m$ is high.
If we used an external constraint, such as from LSS, to keep $\Omega_m$ low,
 the derived $h$ would be higher; this is a result of parameter degeneracies 
in CMB data.

Figure 5 shows the CMB power spectrum corresponding to the 
fitted cosmological parameters and $P_j$'s and also the CMB power 
spectrum for the same cosmological parameters but with all $P_j$'s 
set to unity. One can see where the $P_j$'s are contributing. Most distinctly,
 by allowing the $P_j$'s to vary, the CMB angular power spectrum receives
 significant positive contribution around the first peak.

\begin{figure}[p]
\psfig{file=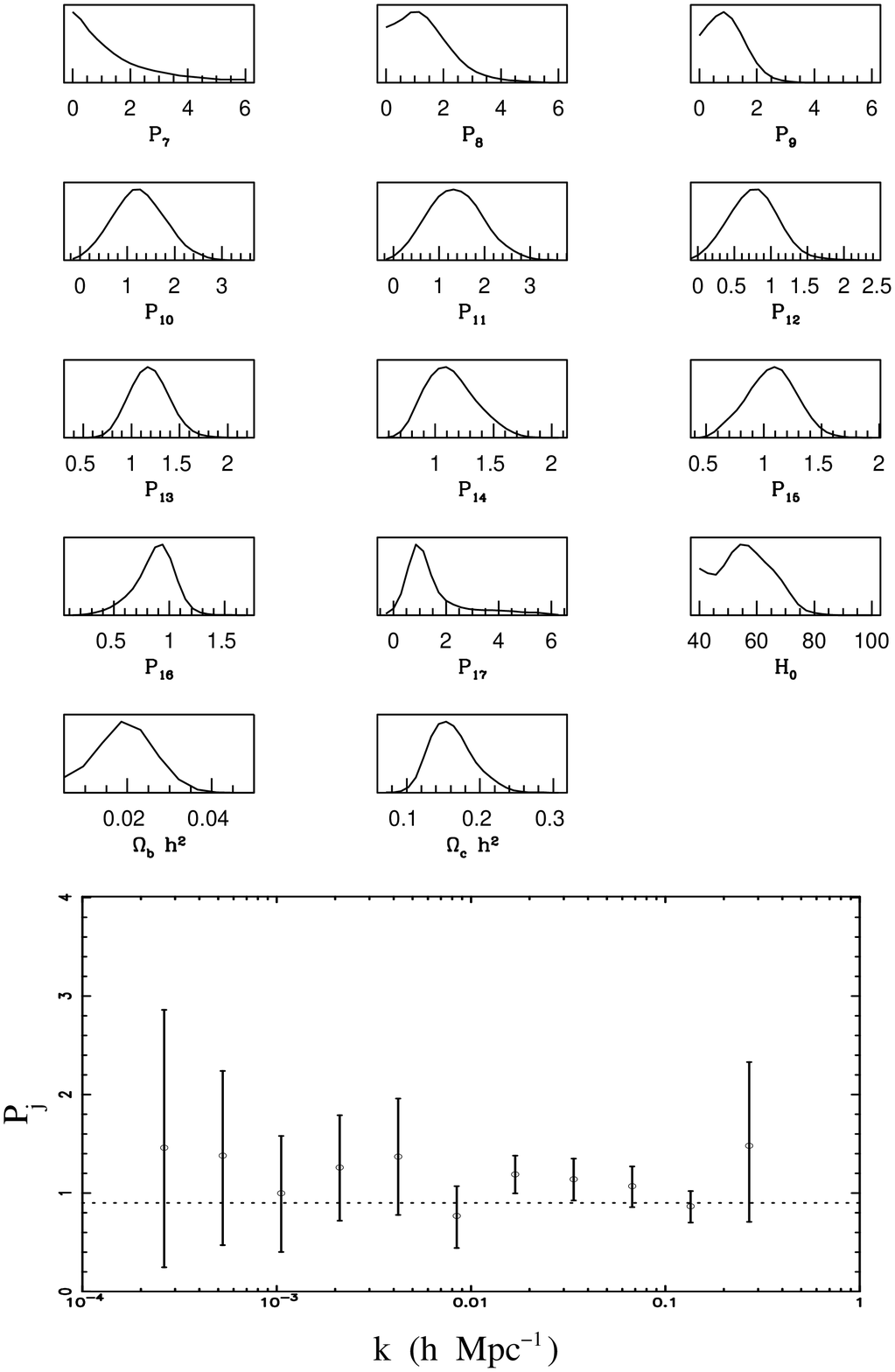,height=7in,angle=0}
\caption{The 1d marginalized posterior distributions obtained upon varying all 
the 14 parameters. The bottom plot shows the constrained $P_j$'s (j=7,17) versus 
scale. The dotted line indicates the best-fit scale-invariant model.
The cosmological parameters are simultaneously constrained to be 
$h=0.56\pm0.09$, $\Omega_b h^2=0.020\pm 0.005$ and $\Omega_c h^2 = 0.161\pm 0.028$.}
\end{figure}

\begin{figure}[p]
\psfig{file=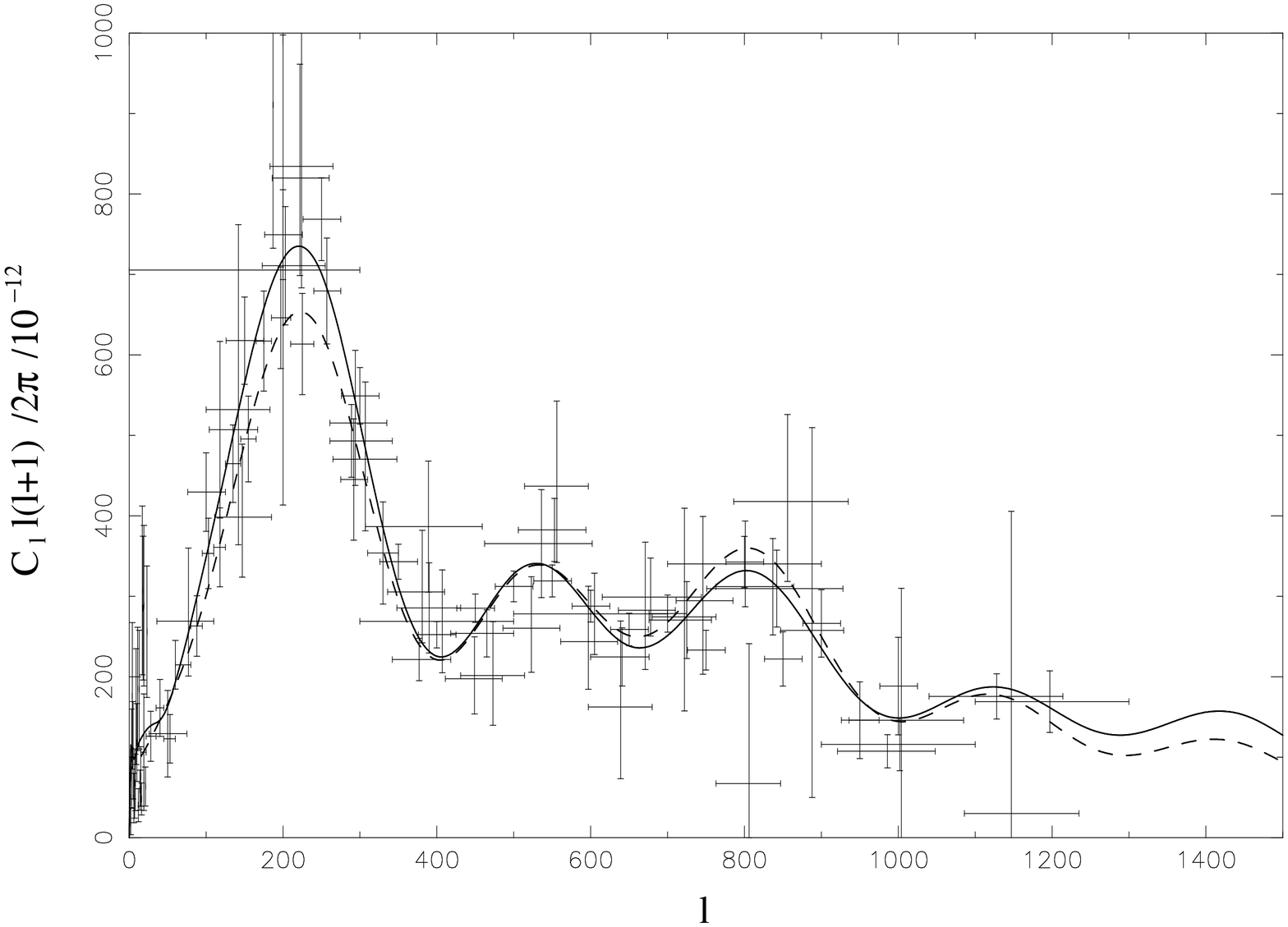,height=3.2in,angle=0}
\caption{The solid line shows the CMB power spectrum at fitted values of 
the cosmological parameters and $P_j$'s.  The dashed line shows the CMB
power spectrum for the same cosmological parameters at $P_j$'s of unity.
  All the data points considered in the analysis are plotted.  The error bars 
do not include calibration and beamwidth uncertainties.}
\end{figure}

The corresponding results for the case of the linear interpolation binning method 
are in Figure 6.
Note that the primordial power spectrum
 reconstructed in this way is consistent with the primordial power spectrum
reconstructed using wavelet band powers [Figure 4].
This is reassuring since in the two methods
 the parameters being reconstructed 
have different correlations amongst themselves and
with the other cosmological parameters [see Figures 2 and 3].
The bin amplitudes $a_1$ to $a_6$ do not rule out the excess power at small
 $k$ that was found by \cite{WangMathews02}.  But from the marginalized
 distributions we see that these bin amplitudes are not  
constrained and vary over the entire range that was allowed in the runs.
In comparison the wavelet banding method works better.
This is as expected since the wavelet band powers are uncorrelated by construction,
hence are less correlated when estimated from data than the linear interpolation bins.

\begin{figure}[p]
\psfig{file=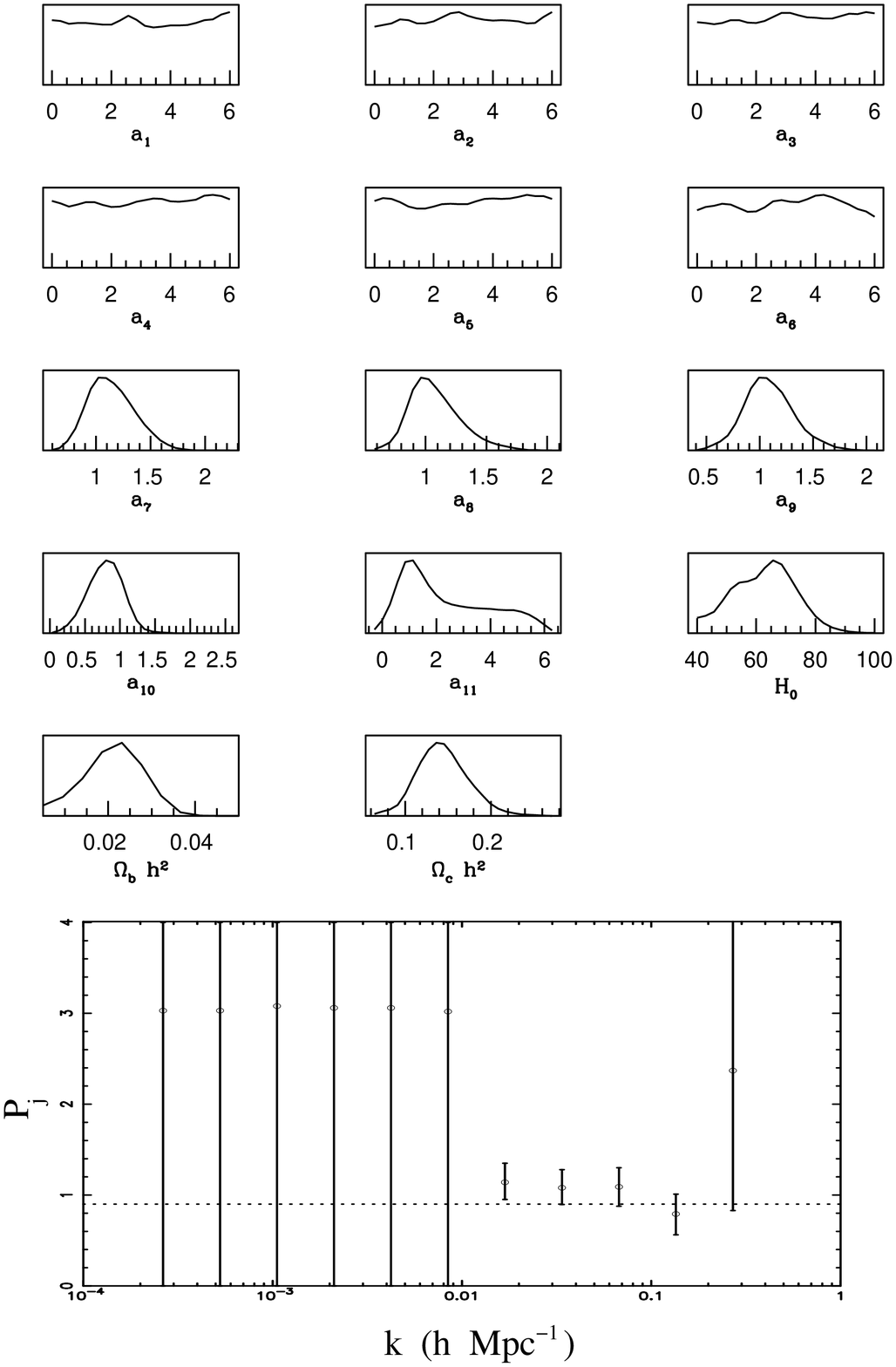,height=7.in,angle=0}
\caption{Results obtained by parameterizing the primordial power spectrum
with bin amplitudes $a_j$s in 11 bins (j=1,11) corresponding exactly to the
 central $k$ values of the wavelet bands. 
 The dotted line indicates the best-fit scale-invariant model.
 The amplitudes $a_1$ to $a_6$ are
 unconstrained; hence their standard deviation is not meaningful. The 
cosmological parameters are simultaneously constrained to be $h=0.63\pm0.10$,
 $\Omega_b h^2=0.021\pm 0.005$ and $\Omega_c h^2 = 0.144\pm 0.029$.}
\end{figure}

\begin{figure}[p]
\psfig{file=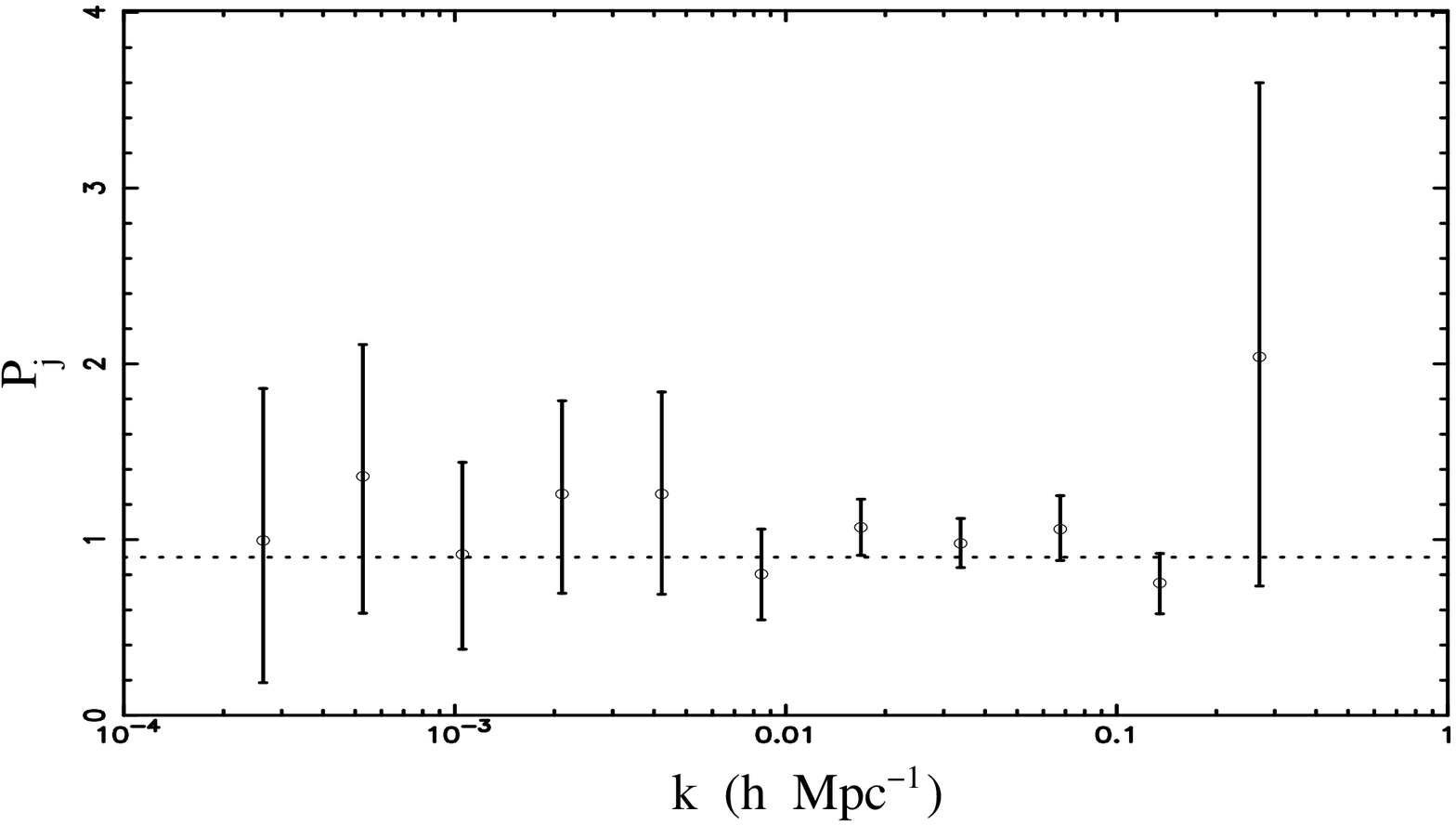,height=3.in,angle=0}
\caption{Same as Figure 4, but with a strong prior on $H_0$ (see text).
 The cosmological parameters are simultaneously constrained to be 
$h=0.64\pm0.62$, $\Omega_b h^2=0.021\pm 0.004$ and $\Omega_c h^2 = 
0.144\pm 0.022$.}
\end{figure}

Now we examine the effect, on the wavelet band power results, of
 imposing statistical priors.
If a Gaussian prior is introduced in the value of $h$, 
a weak conservative prior 
of $0.6\pm0.1$ \citep{Branch98}, together with the above mentioned weak 
age prior, what results is almost identical to 
Figure 4. If a stronger prior on $h$ is introduced, that of 
$h=0.72\pm0.08$ \citep{Freedman01} we find that the feature in the 
$ 0.008 \la k/(h\,\mbox{Mpc}^{-1} ) \la 0.1$ range remains, though it
reduces in amplitude as a whole by nearly 10\% (Figure 7). 
Cosmological parameters are 
simultaneously constrained to be $h=0.64\pm0.06$, 
$\Omega_b h^2=0.023\pm 0.004$ and $\Omega_c h^2 = 0.147\pm 0.022$.
Note that when the $P_j$'s were varied near the currently favored values 
of the cosmological parameters ($h=0.68$, $\Omega_b h^2 = 0.024$, 
and $\Omega_c h^2= 0.12$), the same form for the 
primordial power spectrum, with the whole spectrum shifted down by about 20\%, 
was obtained. 
Without priors on the Hubble constant, 
the current CMB data seem to prefer a somewhat low Hubble constant
of $h=0.56\pm 0.09$ with the freedom that is allowed in
the shape of the primordial power spectrum.
It is not clear whether this is due to any systematic effects.
Note that however, this lowish value of the Hubble constant is
within the uncertainties of other independent measurements
of $H_0$ (for example, see \cite{Branch98,Freedman01,Gott01,Saha01}).

We note that several other groups have studied the constraint
on specific types of features of 
the primordial power spectrum from current observational data;
our findings are consistent with these results
\citep{Silk00,Atrio01,Barriga01,Hannestad01b,Elgaroy02}.
                   
The advantage of our approach in this paper versus previous work is as
follows. We parametrize the primordial power spectrum by its wavelet 
band powers $P_j$. This is a model independent parametrization
of $P_{in}(k)$
\footnote{Miller et al. (2002) do a model independent analysis of the 
CMB angular power spectrum $C_l$'s.}. \cite{Wang99} parametrize the 
primordial power spectrum by a step function [i.e., top-hat banding]; 
this leads to an
estimated primordial power spectrum $P_{in}(k)$ that is {\it discontinuous}, 
while
the $P_{in}(k)$ estimated using wavelet band powers is {\it continuous}.
\cite{WangMathews02} used linear interpolation of
$P_{in}(k)$ values at several $k$ values [equally spaced in log$k$];
this lead to a continuous estimated $P_{in}(k)$, but the $P_{in}(k)$
values estimated from data are strongly correlated. The wavelet band powers
$P_j$ are mutually uncorrelated by construction 
[see Eqs.(\ref{eq:Pj})(\ref{eq:Pin(k)}), though some correlations 
are inevitably introduced because the data
 that are used to constrain them are bands in $l$ space], and 
the banding in this method is not arbitrary.
Note also that by computing the wavelet projections of $C_l$ 
[see Eq.(\ref{Clwaveletproj})] for each set of cosmological parameters
excluding the $P_j$'s, we avoid computing $C_l$ when we vary the wavelet 
band powers $P_j$ in the likelihood analysis. This, together with the MCMC 
technique, made it possible for us to estimate all
the 11 relevant wavelet band powers from the current CMB data
 in a timely fashion.
We expect that our method will be efficient in yielding tighter and detailed
constraints on $P_{in}(k)$ when applied to MAP and Planck data.

Finally we note that the primordial power spectrum results presented here 
are for $k$ in units of $h$ Mpc$^{-1}$. These units are found to yield more 
constraining results and this is important for pre-WMAP data. Some 
published results are for $P_{in}(k)$ with $k$ in units of Mpc$^{-1}$. 
Note that when $P_{in}(k)$ is given with $k$ in units of Mpc$^{-1}$,
the feature we have found in this paper would be
shifted to smaller $k$, to near $k \sim 0.01$ Mpc$^{-1}$.

\section{Conclusions}
A model independent 
determination of the $P_{in}(k)$ could uniquely 
constrain unknown physics in the very early universe, test what we have  
assumed about early universe physics, and provide powerful 
constraints on inflationary models. 
Thus rather than assuming specific forms for $P_{in}(k)$, we 
 have used a wavelet band power expansion to  
extract $P_{in}(k)$ as a free function, using recent high precision
CMB data. 
The wavelet band powers parametrization of the primordial power spectrum 
has the following features: in this scheme the banding is not arbitrary but 
well defined and adaptive. In terms of these band powers the primordial 
power spectrum can be reconstructed as a smooth function. The band powers 
are mutually 
uncorrelated by construction, and they are excellent approximations
of the primordial power spectrum at the central $k$ value of the
wavelet window functions.  Although in estimating these from CMB data, which are 
band powers in multipole space, some correlations are inevitably introduced 
because of the cosmological model dependent nonlinear mapping between 
wavenumber $k$ and multipole $l$ spaces. 

The wavelet band powers of $P_{in}(k)$ that we have extracted
from current CMB data seem to indicate a feature in the primordial power 
spectrum at $ 0.008 \la k/(h\,\mbox{Mpc}^{-1} ) \la 0.1$, though only at 
low significance. A chi square analysis indicates that a model with the 
estimated $P_j$'s and cosmological parameters and the best fit scale 
invariant model
fare about the same. Future data will better help distinguish between 
these models.
The linear interpolation binning approach of \cite{WangMathews02}
 yields an estimated $P_{in}(k)$ with a similar
feature at roughly the same location in $k$ with comparable significance.
Our results are consistent with previous work by \cite{WangMathews02}.
MAP and Planck\footnote{http://astro.estec.esa.nl/Planck/}
data should allow us to put 
a tighter constraint on the primordial power spectrum 
\citep{Lesgourgues99,Wang99,Hannestad01a,Matsumiya02,Tegmark02}.

\acknowledgements
We acknowledge the use of CAMB and CosmoMC. This work was supported by 
NSF CAREER grant AST-0094335 at the Univ. of Oklahoma. We thank the 
referee for helpful comments.

\newpage
\appendix

\section{Wavelet Band Powers}

Here we describe our wavelet parametrization of the primordial power spectrum
in detail.

The wavelet transform bases are obtained from dilations and translations of 
a certain (mother) function $\psi(x)$  via
\begin{eqnarray}
\psi_{j,l}(x) & = & \left(\frac{2^j}{L}\right)^{1/2}\psi(2^jx/L-l).
\label{psibasis}
\end{eqnarray}
where $\psi(x)$ is in general real, defined on the interval $[0,1]$, and obeys 
several restrictive mathematical relations first derived by Daubechies (1992) 
in order for the resulting wavelet basis to be discrete, orthogonal and 
compactly-supported. These are the kind of wavelets we consider here. 
See, for example, Press et al. (1994) for an introduction to wavelets,
 and Barreiro et al. (2000) and Tenorio et al. (1999) for applications of
 spherical wavelets to CMB data on the sky.
The $j$ and $l$ are scale and position indices respectively, and the wavelet bases 
are orthogonal with respect to both these indices,
\begin{eqnarray*}
\int_{-\infty}^{\infty} \psi_{j,l}(x)\psi_{j',l'}(x)\,{\rm d}x  & = &
\delta_{jj'}\delta_{ll'}.
\end{eqnarray*}
A periodic function $f(x)$ of period $L$, sampled at $N=2^J$ equally spaced points 
between $0$ and $L$, can be expanded in terms of the wavelet basis as
\begin{equation}
f(x_i) = \sum_{j=0}^{J-1}\sum_{l=0}^{2^j-1} b_{j,l}\psi_{j,l}(x_i),
\label{wtdig}
\end{equation}
where the coefficients $b_{j,l}$ are given by 
\begin{equation}
b_{j,l} =\int_0^L f(x) \psi_{j,l}(x)\,{\rm d}x.
\label{bcoeffs}
\end{equation}
The scale index $j$ increases from 0 to $J-1$, and the wavelets with increasing 
$j$ represent the structure in the function on
increasingly smaller scales, with each scale a factor of 2 finer than
the previous one.  The index $l$ (which runs from 0 to $2^j-1$)
denotes the position of the wavelet $\psi_{j,l}$ within the $j$th
scale.  Thus $b_{j,l}$ measures the signal in $f(x)$ on scale $L/2^j$, 
and centered at position $lL/2^j$ in physical space and centered at wavenumber 
$2\pi \times 2^j/L$ in Fourier space. 

The Fourier decomposition of function $f(x_i)$ is given by 
\begin{equation}
f(x_i)=\sum_{n=0}^{N-1} \epsilon_n e^{i2\pi nx_i/L}
\label{ft}
\end{equation}
and the Fourier coefficient $\epsilon_n$ is
\begin{equation}
\epsilon_n = \frac{1}{L} \int_0^L f(x) e^{-i2\pi nx/L} dx.
\label{invft}
\end{equation}
Since both the discrete wavelet transform (DWT) and the Fourier transform 
(FT) bases are complete, there exists a relationship between the Fourier and 
wavelet coefficients. Substituting equation (\ref{ft}) in (\ref{bcoeffs}) gives
\begin{equation}
b_{j,l} = \sum_{n=-\infty}^{\infty} \epsilon_n \hat{\psi}_{j,l}(-n),
\label{wtft}
\end{equation}
and similarly equations (\ref{invft}) and (\ref{wtdig}) give
\begin{equation}
\epsilon_n=\frac{1}{L}\sum_{j=0}^{\infty} \sum_{l=0}^{2^j-1} b_{j,l} 
\hat{\psi}_{j,l}(n),
\label{ftwt}
\end{equation}
where $\hat{\psi}_{j,l}(n)$ is the FT of the wavelet $\psi_{j,l}$ 
and is related to the FT of the basic wavelet (using (\ref{psibasis})) by
\begin{equation}
\hat{\psi}_{j,l}(n) = \int_0^L \psi_{j,l}(x)e^{-i2\pi nx/L} dx = 
\left(\frac{2^j}{L}\right)^{-1/2} \hat{\psi}\left(\frac{n}{2^j}\right) 
e^{-i2\pi nl/{2^j}}.
\label{ftofwavelet}
\end{equation}

From (A6), the covariance in wavelet space is given by
\begin{equation}
\langle b_{j,l}b_{j',l'}\rangle=\sum_{n,n'=-\infty}^{\infty} 
\langle\epsilon_n,\epsilon_{n'}\rangle  \hat{\psi}_{j,l}(n) \hat{\psi}^{\dagger}_{j',l'}(n').
\end{equation}
For a homogeneous Gaussian random field, ignoring the often very small correlations 
that may exist between the wavelet coefficients (see  Frazier, Jawerth \& Weiss (1991), 
Walter (1992), Zhang \& Walter (1994), and Tenorio, Stark \& 
Lineweaver (1999); such correlations, ignored in mosts works, reduce further with the 
regularity of the wavelet), $\langle b_{j,l}b_{j',l'}\rangle =P_j \delta_{j,j'} 
\delta_{l,l'}$.\footnote{When the random field is ergodic, 
the $2^j$ coefficients at a given scale can be taken as $2^j$ independent 
measurements. The average over l is thus a fair estimation of the ensemble 
average.}
Also for a Gaussian random field the Fourier amplitudes ($|\epsilon_n|$)
 have a Gaussian one point distribution and their phases are 
random, so that $\langle\epsilon_n, \epsilon_{n'}\rangle = P(n) \delta_{n,n'}$. 
Thus
\begin{equation}
P_j = \frac{1}{2^j} \sum_{n=-\infty}^{\infty} \left| \hat{\psi}
\left(\frac{n}{2^j}\right) \right|^2 P(n).
\label{impeqn1}
\end{equation}
where $P_j$ is the 
variance of $b_{j,l}$, the power of perturbations in wavelet 
coefficients of scale $j$.

The $P_j$'s are thus the scale-by-scale band-averaged Fourier power spectrum, 
from which one can attempt to reconstruct the Fourier power spectrum 
as a smooth function. The weak correlation of the wavelet coefficients
discussed above leads to an excellent approximation
to $P(n)$:
\begin{equation}
\hat{P}(n)= \sum_{j=0}^\infty P_j \left| \hat{\psi}\left(\frac{n}{2^j}
\right)\right|^2.
\label{impeqn2}
\end{equation}
Moreover, for a Gaussian random field the $P_j$'s are (very nearly, see discussion above) 
uncorrelated:
\begin{equation}
\frac{ \langle P_j\, P_{j+1} \rangle}
{P_j \, P_{j+1}}
= \frac{ 2^{j+1} \sum_l b_{j,l/2}^2 \, b_{j+1,l}^2  }
{ \sum_l b_{j,l/2}^2 \, \sum_l b_{j+1,l}^2 } = 1.
\end{equation}
Scale-scale correlations, as defined above for order 2 (the value of the exponent), 
has been discussed in detail by Pando, Valls-Gabaud \& Fang (1998), and 
Mukherjee, Hobson \& Lasenby (2000), and others (note that there are half the number of coefficients at scale $j$ than at scale $j+1$). The existence of detectable scale scale correlations is an indication of mode-mode coupling and hence non-Gaussianity.

In order to parametrize the primordial power spectrum in a model 
independent way, we need to estimate the power in certain bands in $k$. 
The bands should be logarithmically spaced in $k$, as in \cite{WangMathews02}.
To find a unique number for the appropriate number of bands to use,
we adopt the DWT approach to banding. While in the Fourier approach the 
phase space is split such that the resolution in wavenumber $k$ is highest 
at all $k$ $(\Delta k \rightarrow 0)$, and the resolution in position $x$ is 
lowest, $(\Delta x \rightarrow \infty)$, in the wavelet approach these 
resolutions are adaptive.
We choose $\Delta x \propto 1/k$, and $\Delta k/k = \log\,2$,
 so that an optimal chopping of the phase space 
is achieved whilst satisfying the uncertainty relation $\Delta x \Delta k 
\geq 2\pi$ (see \cite{fangfeng} for a discussion). Wavelets afford good 
$k$ resolution at small $k$ and poorer resolution by factors of 2 as $j$ 
(or position resolution) increases.  Further, the wavelet band powers
  $P_j$, for a Gaussian random field, 
are uncorrelated by definition, 
and one cannot have more independent bands 
\citep{fangfeng}. Equations (\ref{impeqn1}) and (\ref{impeqn2}) show how 
the primordial Fourier power spectrum can be parametrized in terms of $P_j$'s 
which represent a scale-by-scale band-averaged Fourier power 
spectrum with log$_{10}(2)$ spacing.

Note the Eq.(\ref{impeqn1}) is essentially just a suitable form of 
banding with window functions shown in Fig. 1. The equations preceding 
it in this section show that if the field in question, here a 
statistically homogeneous and isotropic Gaussian random primordial 
density field, were available to us then the $P_j$'s would be the 
variance of wavelet coefficients of scale $j$. Although we have given 
equations for a 1d field, if the primordial density field is isotropic 
the equivalence can be made.

All results shown in this paper are for the wavelet Daubachies 20. 
We have also studied 
the case for the Symmlet 8 wavelet, obtaining very similar results. The larger 
the number associated with the wavelet, the more smooth is the wavelet in 
real space, and the lesser their compact support in real space (less 
localized, though compact support is technically more involved a concept 
than localization, and while wavelets can be localized in both real and 
Fourier space, it is impossible for a function to have compact support 
in both spaces). Since we hope to be able to pick up sharp 
features in the primordial (Fourier) power spectrum, wavelets that are 
smooth in real space are preferable for our purpose.  In fact 
wavelets that have compact 
support in Fourier space rather than in real space (often called 
band-limited wavelets) should do better. Examples of such wavelets 
are the Shannon wavelet and the Meyer wavelet. We defer their use
 for a future paper as these are less frequently used wavelets 
so that the relevant software is not easily available.

\end{document}